\documentclass[prl,aps,showpacs,twocolumn]{revtex4}

\usepackage{amsmath,amssymb}
\usepackage{graphicx}
\usepackage{psfrag}

\newcommand{\kcomp}{k_{\text{comp}}}
\newcommand{\kbend}{k_{\text{bend}}} 
\newcommand{\xinew}{\xi'}
\newcommand{\munew}{{\mu^\prime}}
\newcommand{\nunew}{{\nu^\prime}}

\newcommand{\orig}[1]{}
\newcommand{\todo}[1]{}

\begin{document}
\title{Elasticity of Stiff Polymer Networks}

\author{Jan Wilhelm and Erwin Frey}

\affiliation{Hahn-Meitner-Institut, Abteilung Theorie, Glienicker
  Strasse 100, D-14109 Berlin, Germany \\ Fachbereich Physik, Freie
  Universit\"at Berlin, Arnimallee 14, D-14195 Berlin, Germany}

 \date{\today}

\begin{abstract}
  We study the elasticity of a two-dimensional random network of rigid
  rods (``Mikado model''). The essential features incorporated into
  the model are the anisotropic elasticity of the rods and the random
  geometry of the network. We show that there are three distinct
  scaling regimes, characterized by two distinct length scales on the
  elastic backbone. In addition to a critical rigidiy percolation
  region and a homogeneously elastic regime we find a novel
  intermediate scaling regime, where the elasticity is dominated by
  bending deformations.
\end{abstract}

\maketitle

The elasticity of cells is governed by the cytoskeleton, a partially
crosslinked network of relatively stiff filaments forming a several
$100$~nm thick shell called the actin cortex~\cite{alberts}. While the
statistical properties of single cytoskeletal filaments are by now
relatively well understood~\cite{wilhelm-frey:96,legoff-etal:02},
theoretical concepts for the elasticity of stiff polymer networks are
still evolving. One major open question is to understand how stresses
and strains are transmitted in such networks. In synthetic gels that
are formed by rather flexible chain molecules the response to
macroscopic external forces is -- on the level of single filaments --
isotropic and entropic in origin.  It is generally believed that
macroscopic stresses are transmitted in such a way that the local
deformations within the network stay affine, i.e. that the end-to-end
distance of individual filaments follows the macroscopic shear
deformation~\cite{doi-edwards:86}. In contrast, the building blocks of
the actin cortex are semiflexible polymers, whose hallmark is an
extremly long persistence length $\ell_p$, which is comparable to the
total contour length $\ell$. As a consequence, the response of such
stiff polymers to external forces shows a pronounced
anisotropy~\cite{kroy-frey:96}. Consider a semiflexible polymer with
one end clamped at a fixed orientation. When forces are applied at the
other end transverse to the tangent vector at the clamped end, the
response may be characterized by a transverse spring coefficient
$k_\perp (\ell) = 3 \kappa / \ell^3$ proportional to the bending
modulus $\kappa$.  Whereas this response is of purely mechanical
origin, the linear response for longitudinal forces is due to the
presence of thermal undulations, which tilt parts of the polymer
contour with respect to the force direction. The corresponding
effective spring coefficient $k_\parallel (\ell) = 6 \kappa^2 / (k_B T
\ell^4)$ is proportional to $\kappa^2/T$ indicating the breakdown of
linear response for very stiff filaments.  In a typical network one
expects the distance between crosslinks $\ell_c$ to be much smaller
than the persistence length and filament length.  Hence we have
$k_\parallel(\ell_c)/k_\perp(\ell_c) = 2 \ell_p/\ell_c \gg 1$, i.e.
the elastic response of the filaments is indeed highly anisotropic.

These anisotropic elastic properties of individual filaments suggests
that the macroscopic elasticty of networks will not only depend on the
number of crosslinks and the density of filaments, but also on the
geometry and architecture of the network. For very regular networks
such as a triangular lattice the longitudinal spring coefficient
$k_\parallel$ will dominate the macroscopic
moduli~\cite{mackintosh-kaes-janmey:95} since the network can not be
strained without a change of the end-to-end distance of individual
polymers.  In other regular network architectures, the softer bending
modes would be dominant~\cite{satcher-dewey:96}.  Naturally, this will
lead to a very different prediction for the elastic modulus of the
network. It is not at all obvious what type of network geometry
(elongation dominated versus bending dominated) is relevant in less
ideal structures with a significant amount of disorder such as in
typical cytoskeletal networks.

\begin{figure}[htb]
   \hfil \includegraphics[width=0.48\columnwidth]{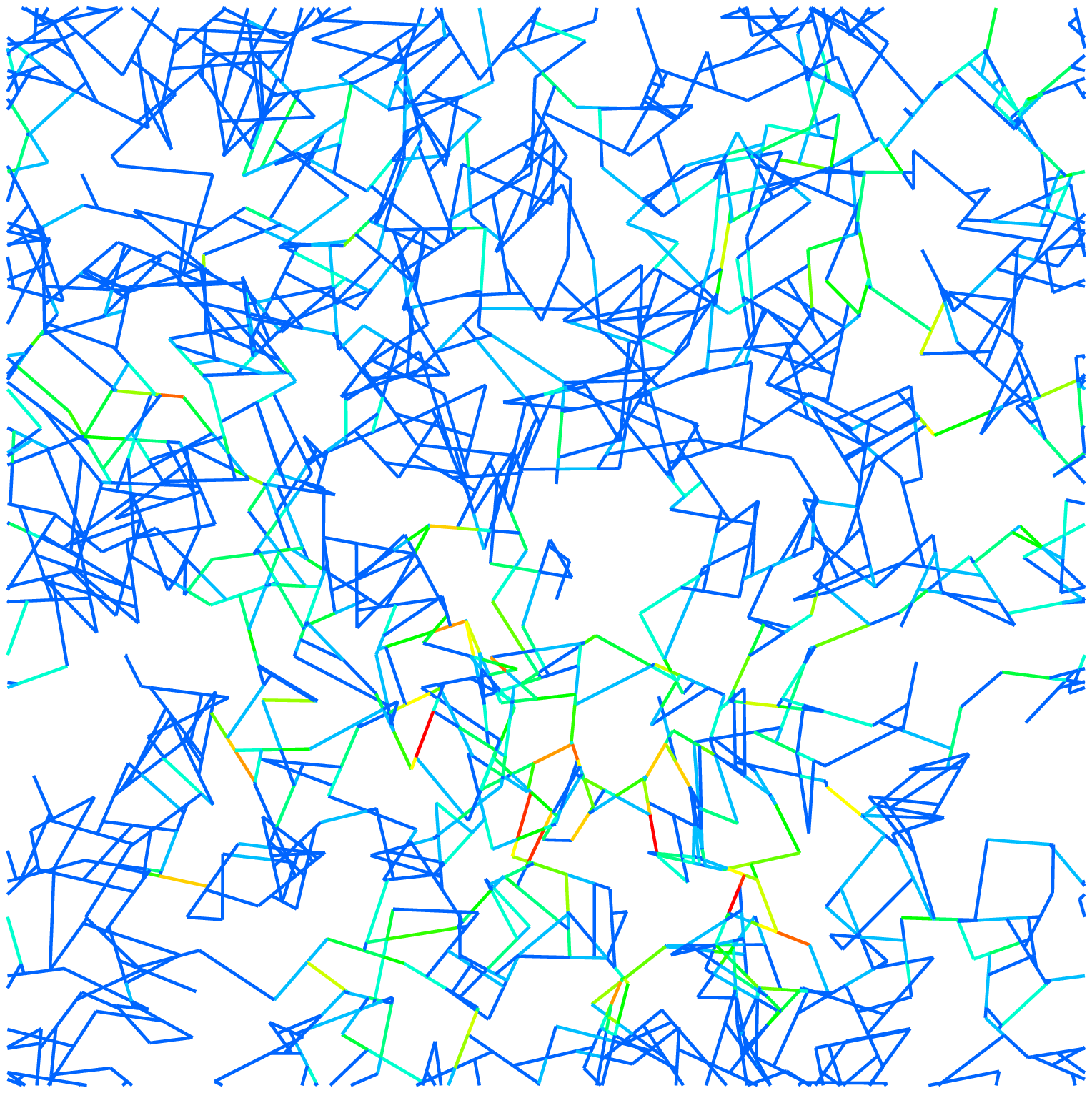} \hfil
   \includegraphics[width=0.48\columnwidth]{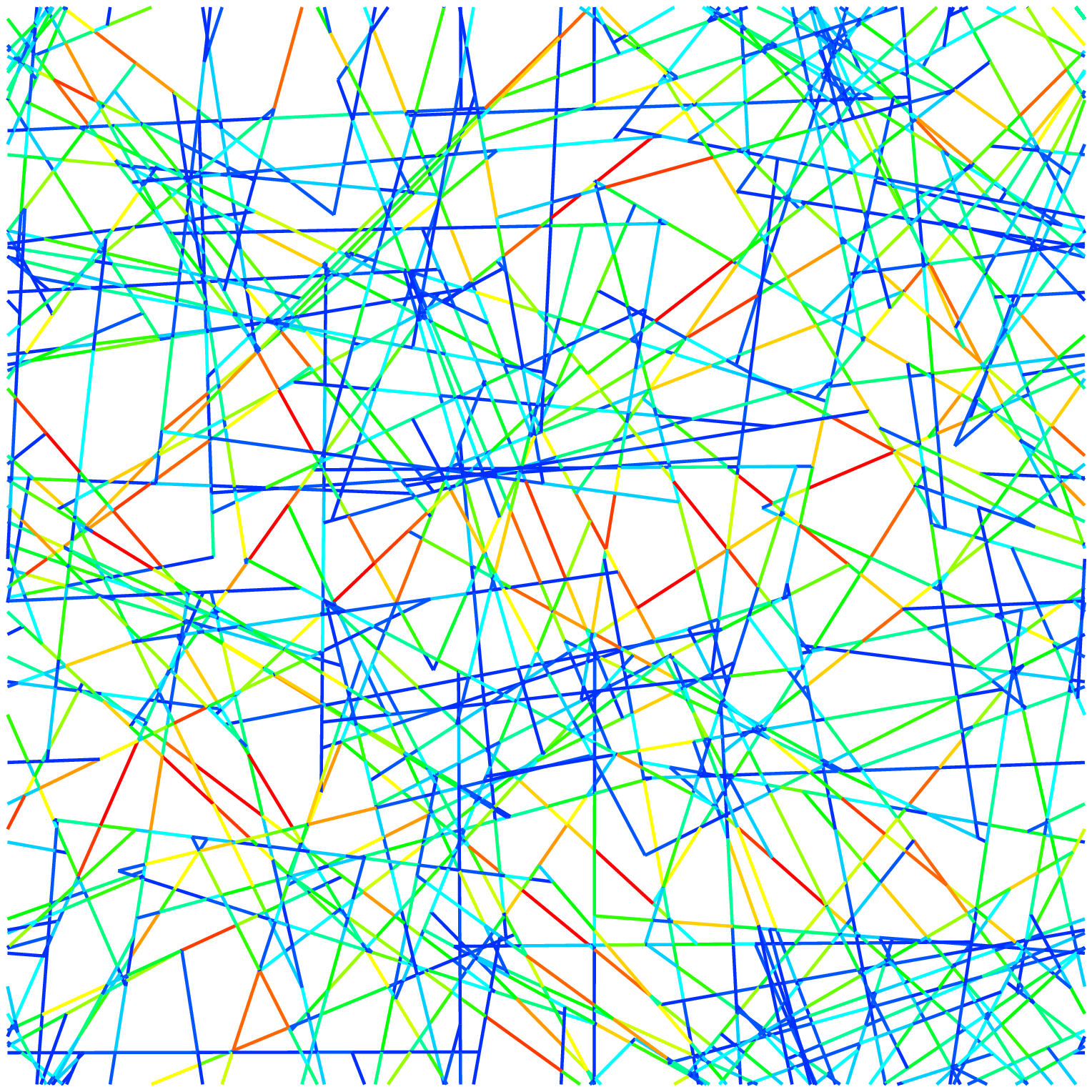} \hfil
\caption{Typical networks at low and high density. Dangling bonds, 
  not contributing to the elasticity, have been cut off. The stress
  distribution is shown in false colors; the load on a filament
  increases from blue to red. The {\em left} picture is for $\rho =
  10$, system size $L = 10$, and an aspect ratio $\alpha = 0.0001$.
  $99.99\%$ of the strain energy stored in bending modes. In contrast,
  the {\em right} picture shows a network for $\rho = 50$, $L = 2$,
  and $\alpha = 0.01$, where only $5\%$ of the strain energy is in
  bending modes; the remainder is stored in compression modes. For the
  choice of units see the main text.}
\label{fig:mikado}
\end{figure}

As a first step towards understanding the elasticity of stiff polymer
networks we consider a two-dimensional model defined as follows (see
Fig.~\ref{fig:mikado}). 
We generate the random network by placing $N$ line-like objects of
equal length $\ell$ on a plane with area $A=L^2$ such that both
position and orientation of the filaments are uniformly randomly
distributed. Periodic boundary conditions in both directions are used.
Upon increasing the line density $\rho = N \ell/A$ there is a critical
threshold $\rho_c$ for geometric percolation~\cite{pike-seager:74a}.
Numerical simulations~\cite{leroyer-pommiers:94} show that the
correlation length $\xi \sim (\rho-\rho_c)^{-\nu}$ of the incipient
infinite percolation cluster scales with a critical exponent $\nu =
4/3$, identical to the value obtained for random site percolation on a
lattice~\cite{stauffer-aharony:94}. Transport of scalar quantities
like the conductivity is also in the same universality class as
lattice models~\cite{balberg-binenbaum-wagner:84}. In order to study
the transport of non-scalar quantities like shear stress we need to
specify how forces are transmitted between the building blocks of the
network. In our {\em ``Mikado model''} the building blocks are
homogeneous elastic rods characterized by a Young modulus $E$ and a
circular cross-section of radius $r$. Wherever two sticks intersect
they are connected by a crosslink with zero extensibility. In the
cytoskeleton one finds a variety of linker proteins with a range of
mechanical properties~\cite{limozin-sackmann:02}. Here we restrict
ourselves to crosslinks that either fix the relative orientation of
the rods (``stiff crosslinks'') or allow free rotation (``free
hinges'').  Similar to thermally fluctuating semiflexible polymers,
the elastic response of a stick segment between two neighboring
crosslinks is characterized by length dependent force constants for
compression or elongation, $\kcomp (\ell_c) = \pi r^2 E/\ell_c$, and
bending, $\kbend (\ell_c) = k_\perp(\ell_c) = (3/4) \pi r^4
E/\ell_c^3$. The distance between two crosslinks $\ell_c$ shows a
Poissonian distribution, where the average distance of crosslinks
along a filament scales as the inverse of the line density,
$\bar{\ell_c}=\pi/\rho$~\cite{pike-seager:74a}.  While this is a
purely mechanical model, it still captures the essential feature that
for typical densities of the network the compressional stiffness is
much larger than the bending stiffness, $\kcomp (\ell_c)/ \kbend
(\ell_c)= (4/3) \ell_c^2/r^2 \gg 1$. It leaves out steric effects due
to thermal fluctuations of the filaments, which give rise to the
plateau modulus in solutions~\cite{hinner-etal:98}.

Consider the energy of the network as a function of the deviations of
the positions of all intersection points and the rod orientations at
the intersection points from their initial values. For small
deformations of the network, this function can be approximated by a
quadratic form that vanishes for vanishing deviations, as---by
construction---the undeformed network is not prestressed.  To analyse
the elastic properties of the model network, a shear deformation
respecting the periodic boundary conditions is enforced by demanding
that corresponding points on the left and right boundary of the
simulation cell undergo equal displacements while the displacements of
corresponding points on the upper and lower boundary of the cell must
agree vertically but differ horizontally by a distance $\Delta =
\gamma L$, where $\gamma$ is the shear strain. The orientation of the
rods at corresponding points on the boundary are required to be equal.
The remaining degrees of freedom are then allowed to relax, i.e. the
harmonic approximation to the energy of the network is minimized in
the presence of the constraints. The derivative of the resulting
energy of the deformed state with respect to the strain $\gamma$ is
proportional to the shear modulus. In pinciple, this reduces the
determination of the shear modulus of a given network to the solution
of a linear equation. However, for interesting parameters (thin rods),
the problem is numerically highly unstable as we are searching for the
lowest point of a complicated high-dimensional valley with extremely
steep slopes but hardly varying base altitude. This problem is best
left to one of the commercially available finite element solvers which
have seen many years of careful optimization and testing. The results
presented below were obtained using the program Nastran by
MSC~Software.

In the following discussion we take the rod length $\ell$ as unit
length and $\kappa/\ell^3$ as unit for the elastic modulus. Then the
independent parameters are the densitiy $\rho$, the system size $L$
and the aspect ratio $\alpha = r/\ell$ of the rods. Note that the
latter is a measure of the relative magnitude of compressional to
bending stiffness.

\begin{figure}[htbp]
\begin{center}
  \includegraphics[width=0.8\columnwidth]{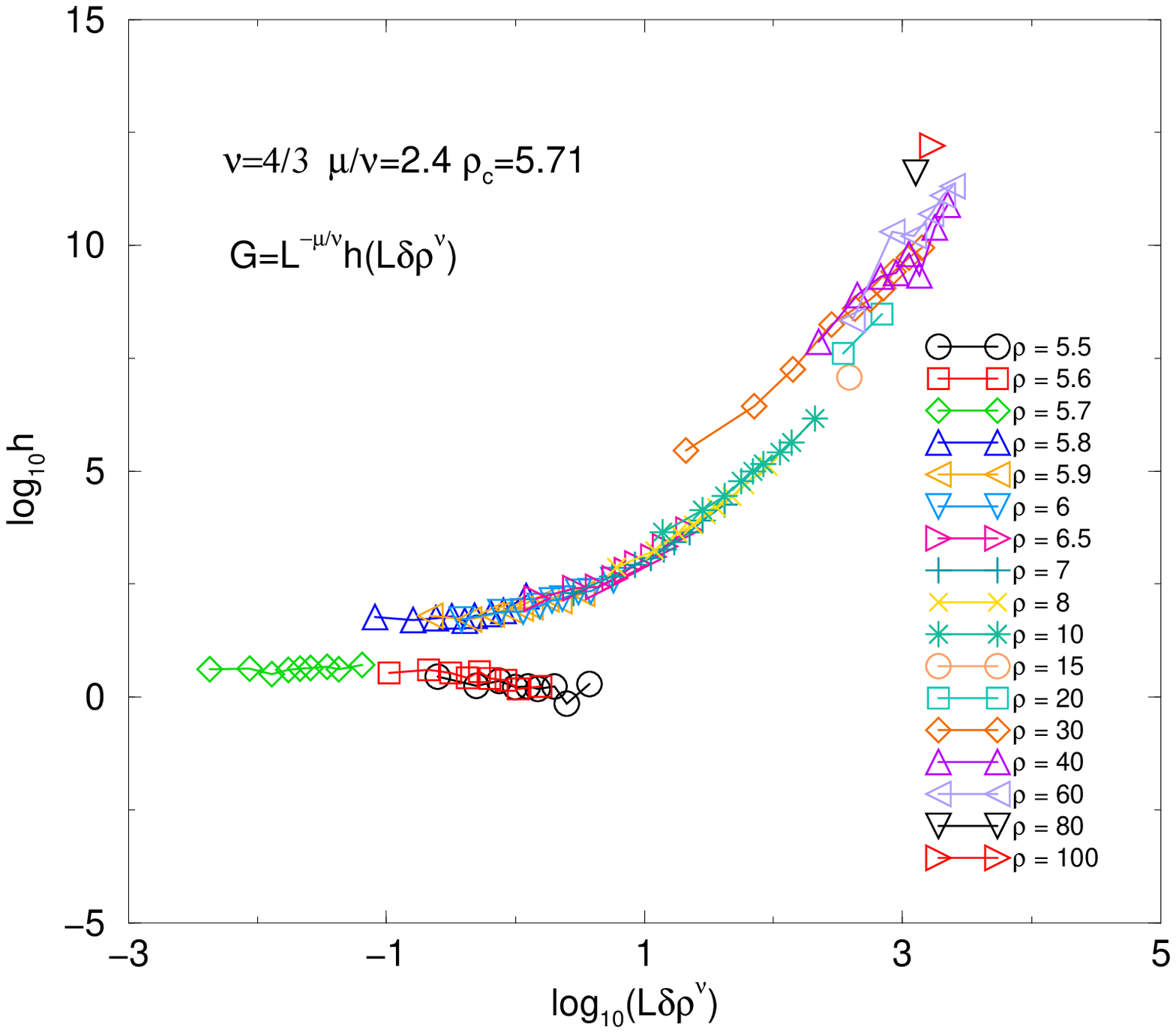} \\
  \includegraphics[width=0.8\columnwidth]{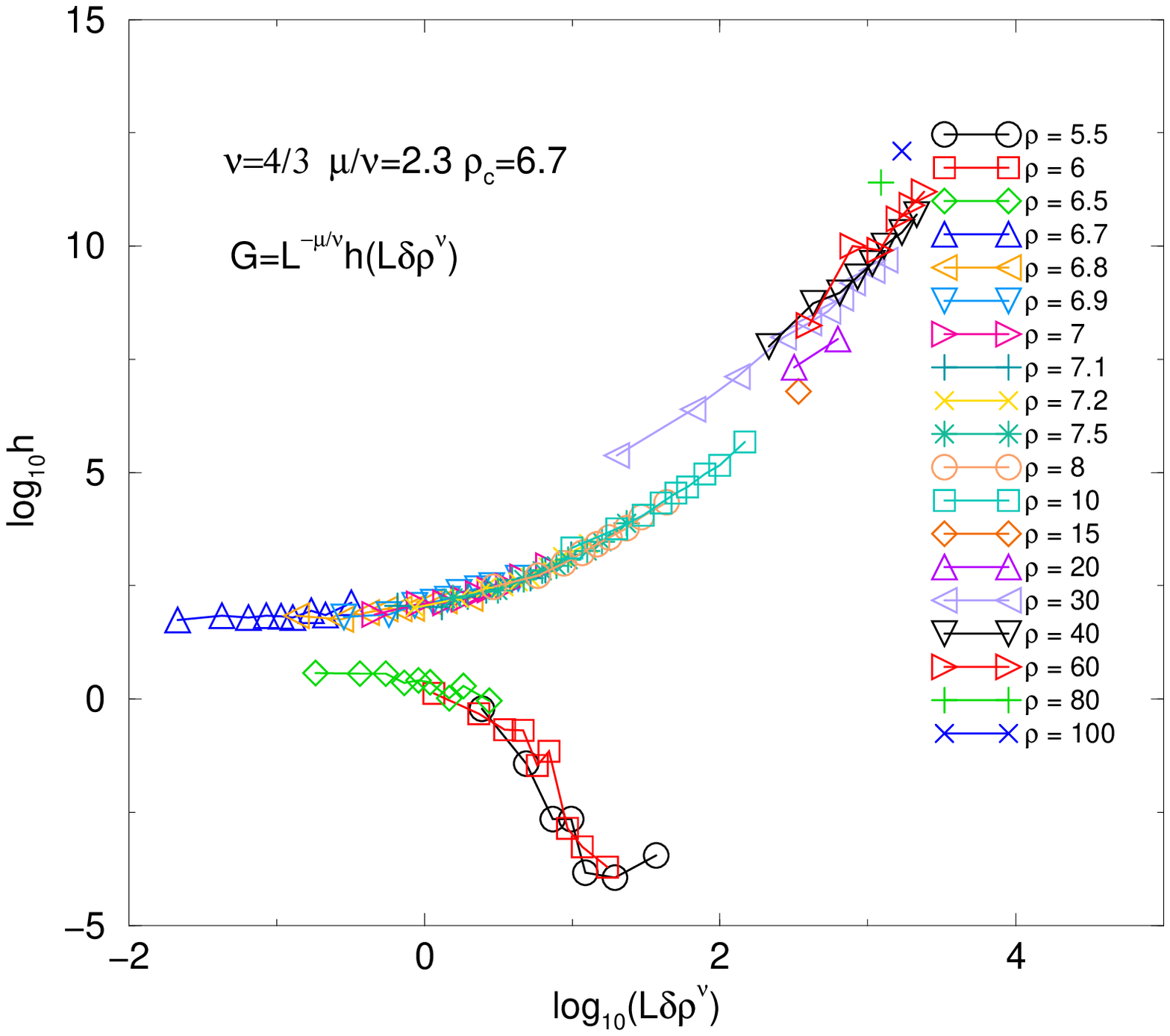}
\end{center}
\caption{Double logarithmic plot of the scaling function $h(x)$ for 
  the shear modulus of the ``Mikado model'' with stiff crosslinks
  (top) and free hinges (bottom) as a function of $x=L |\delta
  \rho|^\nu$ with $\delta \rho = \rho-\rho_c$ for a series of
  densities $\rho$ indicated in the graphs. Note that for finite
  systems the shear modulus is also nonzero below $\rho_c$ (lower
  branches in both plots).}
\label{fig:modul_percolation}
\end{figure}

We start with an analysis of the elasticity close to the percolation
threshold. For stiff crosslinks we find that the percolation threshold
is the same for rigidity as for connectivity percolation, $\rho_c =
5.71$.  For free hinges a higher line density of filaments is needed,
$\rho_c = 6.7$, for the network to become rigid.  This agrees well
with recent results, $\rho_c = 6.68$, for stiff fiber
networks~\cite{latva-kokko-timonen:01}, where the crosslinks are fixed
in space but the angles between the fibers can vary. In both cases, we
find that the shear modulus $G$ vanishes as the line density of sticks
approaches the critical value $\rho_c$, according to a power law $G
\sim (\rho - \rho_c)^\mu$.  For our numerical analysis on a finite
lattice we expect the shear modulus to obey the following finite size
scaling law
\begin{eqnarray}
  G = L^{-\mu/\nu} \, h(L/\xi) \, ,
\label{eq:critical_scaling_law}
\end{eqnarray} 
where the scaling function behaves as $h(x) \sim x^{\mu/\nu}$ and
$h(x) \sim 1 $ for large and small values of the scaling variable
$x=L/\xi$, respectively. Fig.~\ref{fig:modul_percolation} shows that
the data collapse works very well for densities ranging from values
close to the percolation threshold $\rho_c$ up to $\rho
\approx 20$. For the data shown, $L$ ranges from $2$ to $30$.
For larger densities, systematic deviations are clearly visible. This
will turn out to be a very interesting observation, as we will discuss
in detail below.  We get the best data collapse in the critical region
if we choose the values $2.4\pm 0.2$ and $2.3 \pm 0.2$ for the
critical exponent $\mu/\nu$ in the case of stiff crosslinks and free
hinges, respectively.  Since the difference between the exponents is
within the statistical error, we can make no definite conclusion
whether networks with free hinges and stiff crosslinks belong to
different universality classes for elasticity percolation.

The rigidity exponent $\mu \approx 3.15 \pm 0.2$ is significantly
lower than in other classes of continuum percolation models, like the
``Swiss-cheese model'', where $\mu \approx
5$~\cite{feng-halperin-sen:87,benguigui:86}. It is also lower than the
value $\mu \approx 4$ for lattice models with bond-bending forces
\cite{arbabi-sahimi:93b,stauffer-aharony:94}. Hence it seems likely
that the ``Mikado model'' constitutes a new universality class for
rigidity percolation. Similar results have independently been found in
Ref.~\cite{head-levine-mackintosh:03}.

\begin{figure}[htbp]
\begin{center}
  \includegraphics[width=0.8\columnwidth]{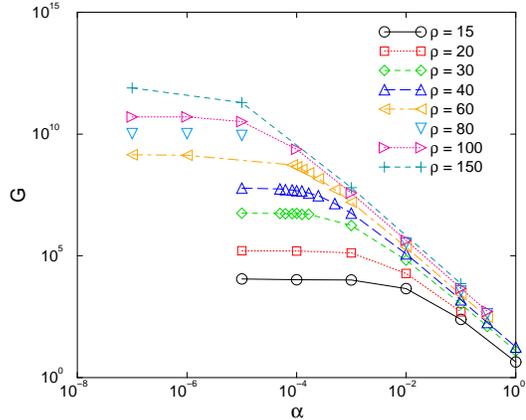}
\end{center}
\caption{Double logartithmic plot of the shear modulus $G$ as a 
  function of $\alpha$ for fixed $\kbend (\ell)$. Data are shown for
  free hinges.}
\label{fig:raw_data}
\end{figure}

\begin{figure}[htbp]
\begin{center}
  \includegraphics[width=0.8\columnwidth]{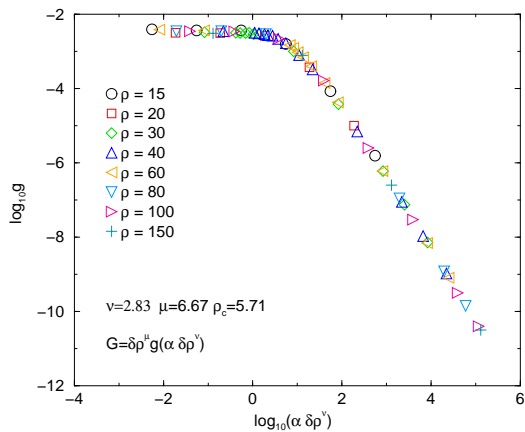}
\end{center}
\caption{Scaling plot of the shear modulus for free hinges
  for a series of densities above $\rho=15$ indicated in the graph
  (same data as in Fig.~\ref{fig:raw_data}). Data collapse to the
  crossover scaling form, Eq.~{\protect\ref{eq:anomalous_scaling}}, is
  obtained with $\nunew = 2.83$. Note that here and in all other
  figures the unit of length is $\ell$ and the unit of the shear
  modulus $G$ is $\kappa / \ell^3$.}
\label{fig:crossover_scaling}
\end{figure}

Now we come back to the above mentioned systematic deviations from the
scaling law, Eq.~\ref{eq:critical_scaling_law}, at densities above
$\rho \approx 20$. To understand these better, let us have a closer
look at the shear modulus as a function of the elastic moduli of
individual filaments for densities not too close to the percolation
threshold.  In this regime the shear modulus becomes independent of
system size for moderately large systems; for the following results we
have chosen systems satisfying $L/\xi \ge 200$.
Fig.~\ref{fig:raw_data} shows the shear modulus as a function of
$\alpha$ for a series of densities; we have communicated a preliminary
version of these data in Ref.~\cite{frey-kroy-wilhelm-sackmann:97}.
Note that $\kbend (\ell)$ is effectively kept constant since we are
measuring all elastic constants in units of $\kappa / \ell^3$.
There are two strikingly different regimes. For high densities and/or
thick rods ($\alpha \agt 0.1$), where compressional stiffness is lower
or comparable to the bending stiffness (lower right part of
Fig.~\ref{fig:raw_data}), the shear modulus scales linearly with the
filament compressional modulus and the number of filaments per unit
area, $G \sim (\rho-\rho_c) \alpha^{-2}$. Such a linear regime has
also been found in a series of studies on random fiber
networks~\cite{raisanen-etal:97}. It is by now well established that
the elastic modulus can be described quantitatively in terms of
effective medium models~\cite{astrom-etal:00}. Hence, in the high
line density regime the network behaves as a homogeneously elastic medium,
dominated by the compressional modulus of the individual filaments.
As a consequence, local deformations follow a macroscopic shear in an
affine way.  This has to be contrasted with the elastic behavior for
slender rods with low aspect ratios ($\alpha \approx 10^{-5}$ for the
higher densities), where bending becomes the softer mode. Then, one
finds an extended plateau region, which broadens significantly with
lowering the line density, where the shear modulus becomes completely
independent of $\kcomp (\ell) \sim \alpha^{-2} \kbend
({\ell})$~\cite{frey-kroy-wilhelm-sackmann:97}.  This strongly
suggests that in this regime the macroscopic elasticity of the network
is dominated by bending stiffness of individual filament.  This
conclusion is corroborated by the observation that almost all of the
energy stored in the deformed network is accounted for by transverse
deformation of the rods (compare Fig.~\ref{fig:mikado}).  Another
remarkable feature of this plateau regime is the strong dependence of
the shear modulus on line density.  We find $G \sim (\rho-\rho_c)^\munew$
with a rather large exponent $\munew \approx 6.7$. From the above
analysis it may seem as if the anomalous elasticity in the plateau
regime and the homogeneous elasticity in the affine regime are two
separate phenomena, and one might wonder how one emerges from the
other. To analyze this relation we try a crossover scaling ansatz,
\begin{equation}
  \label{eq:anomalous_scaling}
   G = (\rho-\rho_c)^\munew g[\alpha (\rho-\rho_c)^\nunew] = 
   \xinew^{-\munew/\nunew} {\tilde g}(\alpha/\xinew) \, ,
\end{equation}   
where we have defined a new length scale $\xinew \sim
(\rho-\rho_c)^{-\nunew}$. For this ansatz to reduce to the modulus
expected in the affine region, the scaling function $g(x)$ needs to
scale as $g(x) \sim x^{-2}$ for $x \gg 1$ {\em and} the exponents need
to obey the scaling relation $\munew = 2\nunew + 1$. In the plateau
regime, $g(x)$ is expected to be constant. As shown in
Fig.~\ref{fig:crossover_scaling}, we obtain an excellent scaling
collapse for over almost eight orders of magnitude in the scaling
variable $x=\alpha/\xinew$ using $\nunew=2.83$ or equivalently $\munew
= 6.67$ and the critical line density $\rho_c \approx 5.71$, associated
with connectivity percolation.  Additionally, the scaling function
$g(x)$ displays the expected behavior. Meeting both of these
requirements is highly nontrivial, and gives strong evidence for the
anomalous scaling law in Eq.~\ref{eq:anomalous_scaling}.

The existence of such a broad scaling regime far from the percolation
threshold is a surprising and intriguing feature of stiff polymer
networks. At the moment we are lacking a complete understanding of its
physical origin. In particular, the geometrical significance of the
new length scale $\xinew$ remains unclear. One may speculate that the
anomalous scaling behavior is a subtle consequence of the interplay
between quenched randomness of the network structure and long-range
correlation effects induced by the stiffness of the filaments. An
immediate consequence of the scaling form,
Eq.~\ref{eq:anomalous_scaling}, is the existence of a crossover line
density $\rho_\text{cross}$ scaling as $ \ell \rho_\text{cross} \sim
\alpha^{-1/\nunew}$, where we have re-introduced units of length
$\ell$. This implies that increasing filament length at constant line
density drives the system towards the affine regime, in accord with
Ref.~\cite{head-levine-mackintosh:03}.

While these results for an idealized two-dimensional model are
certainly not straightforwardly applicable to three-dimensional
cytoskeletal networks, one may still try to get an idea for the scales
involved. We expect that network densities can be compared roughly by
using the average distance $\ell_c$ between intersections as a
measure: A cytoskeletal network might have $\ell_c \approx
0.1\,\mu\text{m}$ with typical filament lengths of $2\,\mu\text{m}$.
These values correspond to a two-dimensional line density of $\rho \approx
20$ and an aspect ratio of $\alpha\approx 0.002$, which would place a
typical actin network in the bending dominated intermediate regime.

Understanding the full complexity of cytoskeletal networks certainly
merits further theoretical and experimental work. Building on the
knowledge gained from our idealized model, future investigations may
among many other questions want to address three-dimensional systems,
polydispersity, thermal fluctuations or even the kinetics of the
crosslinking molecules.

We acknowledge M. Alava and K. Kroy for useful discussions and
comments, and P. Benetatos for a critical reading of the manuscript.

\end{document}